\documentclass[aps,prl,twocolumn,amsmath,amssymb,superscriptaddress,hidelinks]{revtex4-2} 
\usepackage{bm}
\usepackage{verbatim} 

\usepackage{graphicx}
\usepackage[usenames,dvipsnames]{color}
\usepackage[colorlinks]{hyperref}
\hypersetup{
  colorlinks,
  citecolor=blue,
  linkcolor=blue,
  urlcolor=blue}
\usepackage{natbib}

\newcommand{\be}{\begin{equation}}
\newcommand{\ee}{\end{equation}}
\newcommand{\bea}{\begin{eqnarray}}
\newcommand{\eea}{\end{eqnarray}}

\newcommand{\p}{\partial}

\renewcommand{\vec}[1]{{\bf #1}}

\newcommand{\addXY}[1]{\textcolor{blue}{#1}}

\begin{document}

\title{Ultralow p-type contact resistance for ultra-nanoscaled 2D-materials transistors}

\author{Ying Xiong}
\affiliation{Science, Mathematics and Technology, Singapore University of Technology and Design, Singapore 487372}
\author{Tong Su}
\affiliation{Science, Mathematics and Technology, Singapore University of Technology and Design, Singapore 487372}
\affiliation{Zhejiang Province Key Laboratory of Optoelectronic Metrology and Precision Instruments, College of Optical and Electronic Technology, China Jiliang University, 310018 Hangzhou, China}
\author{Qiang Li}
\affiliation{Science, Mathematics and Technology, Singapore University of Technology and Design, Singapore 487372}
\affiliation{Department of Physics, Hubei Minzu University, Enshi, 445000, P. R. China}
\author{Yee Sin Ang}
\affiliation{Science, Mathematics and Technology, Singapore University of Technology and Design, Singapore 487372}
\author{Lain-Jong Li}
\affiliation{Department of 
Materials Science and Engineering, National University of Singapore, Singapore, Singapore}
\author{L. K. Ang}
\email{ricky\_ang@sutd.edu.sg}
\affiliation{Science, Mathematics and Technology, Singapore University of Technology and Design, Singapore 487372}

\begin{abstract}
High contact resistance is one of the main bottlenecks for practical two-dimensional (2D) materials transistors, especially for p-type transistors and future 2D ultra-nanoscaled (sub-10 nm) FETs (PMOS + CMOS).
We develop self-consistent contact resistance models for metal-2D semiconductor-metal devices to capture the essential interface physics for both vertical and edge configurations.
Our calculations have been verified with various recent experiments of p-type and n-type contacts. 
For a given set of materials, the model determines the scaling of contact resistance over a wide range of device parameters including channel length (100s nm down to sub-10 nm), doping and mobility of the 2D materials, contact length of the electrodes, and applied voltages. 
These results identify the key factors in order to reduce the contact resistance for p-type 2D semiconductor WSe$_2$ towards the sub-10 nm channel length scale that are readily to be realized by future experiments.
It is found that the effect of source-limited current saturation is the key challenge for down scaling 2D FET to sub-10 nm channel length.
Two topological semi-metals as potential electrodes are proposed for 2D p-type semiconducting WSe$_2$ with our predicted contact resistance $R_c<$ 100 $\Omega \; {\rm \mu m}$ approaching the quantum limit. 
Our model is also verified with the computational expensive full quantum atomistic model that is currently limited to a few nm scale.
\end{abstract}

\maketitle

2D semiconductors such as transition metal dichalcogenides are promising channel materials for continued down-scaling of future low-power high-performance field effect transistors (FETs) due to their high mobility, clean atomically flat interface and lack of major short channel effect \cite{Akinwande2019,Kim2024,OBrien2023,Das2021,Zeng2024, Lemme2025, Yi2024}. However, contact resistance ($R_c$) is one of the main hurdles for practical 2D FETs according to the International Roadmap for Devices and Systems (IRDS) roadmap especially for sub-10 nm channel length scale.
Although significant progress has been made to achieve ultralow contact resistance ($R_c<$ 100 $\Omega \; {\rm \mu m}$) for n-type 2D FET with MoS$_2$ as the channel material by reverse sputtering~\cite{Fa2025}, employing metallic buffer~\cite{Jiang2024} and using semimetals as contacts \cite{Shen2021,Wu2026,Li2023,Yang2026}, reducing $R_c$ for p-type 2D semiconductors like WSe$_2$~\cite{Zeng2024,Das2026,Kim2024-WSe2,Hoang2026-Pd-WSe2,Das2026-MoSe2-WSe2,Zhang2026-SnS-WSe2,Sun2026-NbWSe2,Wang2026-Se-WSe2,Lance2026Science,Sun2026, Zhao2025} to be comparable to that of n-type remains an ongoing challenge.

\begin{figure*}
  \includegraphics[width=0.96\textwidth]{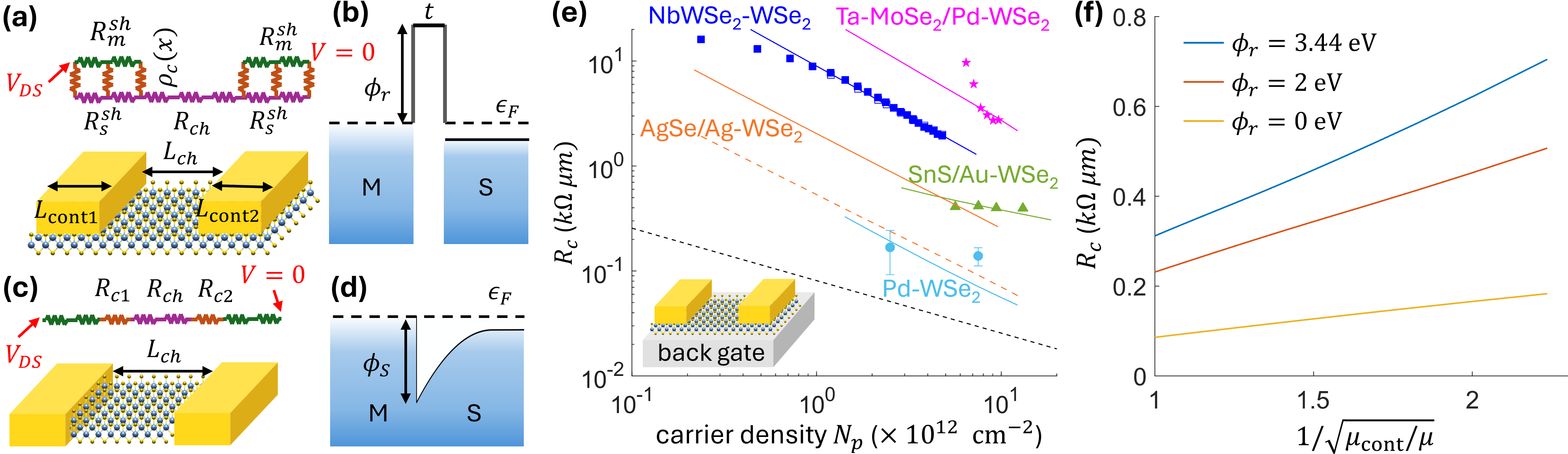}
  \caption{{\bf Modeling contact resistance for metal-2D semiconductors.} (a, c) Schematic diagrams of vertical (a) and edge (c) contacts; the contact resistance is governed by vdW barrier (b) and Schottky barrier (d) respectively. (e) $R_c$ as a function of $N_p$ for the vertical contact. Symbols denote experimental data, and solid lines denote theoretical models. $\eta = 0$ ($k_{tr}$ conserving) produces calculations for NbWSe$_2$-WSe$_2$~\cite{Sun2026-NbWSe2} ($\phi_r=1.65$ eV and $t=0.3526$ nm) and Ta-MoSe$_2$/Pd-WSe$_2$~\cite{Das2026-MoSe2-WSe2} ($\phi_r=1.1$ eV, and $t=0.376$ nm). Here, $t$ is extracted from DFT calculations as the interlayer distance for AA stacking. $\eta = 1$ ($k_{tr}$ nonconserving) is used for SnS/Au-WSe$_2$~\cite{Zhang2026-SnS-WSe2} (with $\phi_r$=0, $t=1$) and for Pd-WSe$_2$~\cite{Lance2026Science} (with $\phi_r =2.36$ eV and $t=0.23$ nm). New proposed contact (AgSe/Ag-WSe$_2$) is simulated with $\eta = 0$, with $\phi_r=3.44$ eV, $t=0.137$ nm (orange solid) and $\phi_r=0$, $t=0$ (orange dashed). The black dashed line indicates the quantum limit. We have set $\mu_{\rm cont} = \mu$ in (e).  
  (f) The effect of $\mu_{\rm cont}/\mu$ on $R_c$ for Aa/AgSe-WSe$_2$ is shown with different barrier heights for fixed $t=0.137$ nm. See {\bf SI} for details. 
}
  \label{Fig1}
\end{figure*}

For proper design of low contact resistance for both n-type and p-type, physics-consistent models for the contact resistance of metal-2D semiconductors-metal (MSM) configuration are lacking over a broad range of length scale from 100s nm down to sub-10 nm for both vertical and lateral contact configurations.
First-principles and non-equilibrium Green's function calculations are computationally expensive and not possible for length scale longer than a few nanometers.
For classical silicon-based electronics, the contact is ohmic where the charge transport is mainly based on thermionic emission known as the Schottky diode equation, which requires revision for 2D materials \cite{Banerjee2020,Ang2016,Ang2018,Ang2019}.
For 2D materials based electrical contact, electron tunneling across a finite interface barrier must be included.
For ultra-nanoscaled channel between the two electrodes, which can be at sub-10 nm scale, the physics of the entire MSM device must be included.
Thus, a consistent device model to determine the MSM contact resistance is necessary to design an ultralow p-contact approaching the industry requirement similar to the n-type contact resistance accomplished experimentally a few years ago.

We consider two types of MSM configurations: vertical contact (Fig.~\ref{Fig1}a, b) where metal forms a van der Waals (vdW) 2D interface with the semiconductor~\cite{Pourfath2026}, and edge contact (Fig.~\ref{Fig1}c, d) where a Schottky junction is formed. 
We develop generic models for both configurations which self-consistently solve the current transport $J$ and contact resistance $R_c$ of the entire MSM device. We analyze the scaling of the contact resistance with respect to experimentally relevant parameters and provide guidelines for further lowering $R_c$ towards the quantum limit.
We have benchmarked our models with recent experimental results for p-contact (with WSe$_2$ as channel material) and predict its expected outcome when the channel length is reduced to sub-10 nm scale. 
In addition to the known contact materials (or electrodes) used in these experiments, we also propose new contact materials to further reduce the p-type contact resistance. 
While we focus on the p-contact problems in this paper, our models have been adapted to n-contact with good agreement with many experiments (with MoS$_2$ channel) as discussed in Appendix A.


\addXY{\it Contact resistance for vertical and edge metal-2D semiconductor contact.} 
We model the current across the contact by accounting for the tunneling of an electron from metal with Bloch wavevector $\vec k$ and  energy $\epsilon_{n, \vec k}$ to the semiconductor at $\vec q $ and energy $\epsilon_{\vec q} = \epsilon_{\rm VBM} +\hbar^2 q^2/2m_s$ along the injection direction $k_{\rm inj}$. The tunneling current at vertical ($v$) and edge ($e$) contacts can be written as 
\begin{align}\label{eq:Jv}
J_{v,e} (V) = \frac{ge}{(2\pi)^d} \sum_n &\int d\vec k  \frac{1}{\hbar} \frac{\p \epsilon_{n, \vec k}}{\p k_{\rm inj}} \mathcal T (\epsilon_{n, \vec k}, V) \Delta f,
\end{align}
where $d=3$ for vertical contact with a bulk metal, $d=2$ for edge contact, $\Delta f = f (\epsilon_{n, \vec k}, \epsilon_F) - f (\epsilon_{n, \vec k}, \epsilon_F + V)$ is the difference between the electron distribution functions in metal and semiconductor, and $f (\epsilon, \epsilon_F ) = 1/[1+ e^{(\epsilon-\epsilon_F)/k_B T}]$. Here $V$ is defined as the voltage drop from the metal to the semiconductor. For edge contact, The tunneling probability $\mathcal T$ is constrained by conservation of energy and momentum transverse to the current direction $\vec k_{\rm tr} = \vec q_{\rm tr}$. For vertical contact, momentum conservation may be relaxed due to interfacial scattering effects~\cite{meshkov1986tunneling,chandni2016}. Thus the tunneling probability becomes 
\be\label{eq:T_final}
\mathcal T (\epsilon_{n, \vec k}, V) = T_{v,e} (\epsilon_{n, \vec k}) \big[\eta + (1-\eta) \Theta(\epsilon^s_{\rm VBM} + V - \epsilon_{\rm inj}^s) \big],
\ee
where $T_{v,e} (\epsilon_{n, \vec k})$ is derived from the Wentzel-Kramers-Brillouin approximation \cite{bohm1951quantum} (see later), $\eta = 0$ for $\vec k_{tr}$ conserved, $\eta = 1$ for $
\vec k_{tr}$ not conserved, $\epsilon_{\rm inj}^s = \epsilon_{n,\vec k} - \epsilon_{\rm tr}^s$ and $ \epsilon_{\rm tr}^s$ are the energies associated with the electron injection direction and transverse direction respectively. For injected electrons from metallic contact with isotropic parabolic dispersion $\epsilon_{n,\vec k} = \epsilon_{n}^0 + \frac{\hbar^2 k^2}{2m_m}$, we have $\epsilon_{\rm tr}^s = \frac{m_m}{m_s} (\epsilon_{n,\vec k} - \epsilon_{n}^0 )\sin^2 \theta $ due to conservation of transverse momentum, where $\theta = \tan^{-1}(k_{\rm tr}/k_{\rm inj})$. See Supplementary Information ({\bf SI}) for the derivation of $\epsilon_{\rm tr}^s$ for generic metallic bands. 

For vertical contact, tunneling via the vdW barrier is approximated by a rectangular barrier (Fig.~\ref{Fig1}b) with height $\phi_r$ and width $t$:
\be\label{eq:Tv}
T_v (\epsilon) = \min \left\{\exp\left[ -\frac{2t \sqrt{2m_0}}{\hbar} \sqrt{\phi_r -\epsilon_{\rm inj}^{\rm vac} }\right], 1 \right\}.
\ee
Here, $m_0$ is the electron mass in vacuum, and $\epsilon_{\rm inj}^{\rm vac} = \epsilon_{n,\vec k} - \epsilon_{\rm tr}^{\rm vac}$ and $\epsilon_{\rm tr}^{\rm vac} = \epsilon_{\rm tr}^{s} m_s/m_0$ are the kinetic energies associated with the electron injection direction and transverse direction in the vacuum. 

The contact resistance of a vertical contact depicted in Fig.~\ref{Fig1}a can be described by a generalized transmission line model (TLM), where the current at each point along the interface is governed by electron tunneling through Eq.~(\ref{eq:Jv}). The total current flowing through the contact must be self-consistently calculated by solving the following TLM governed equations \cite{banerjee2019TLM}: 
\begin{align}\label{eq:TLM}
    &\frac{\p J_{m,x} }{\p x} = -J_v(V_{m,x}-V_{s,x}), \quad \frac{\p V_{m,x}}{\p x} = -J_{m,x} R_m^{sh}, \nonumber \\
    &\frac{\p J_{s,x}}{\p x} = J_v(V_{m,x}-V_{s,x}), \quad \quad \frac{\p V_{s,x}}{\p x} = -J_{s,x} R_s^{sh},
\end{align}
where $V_{m(s),x}$ and $J_{m(s),x}$ are the voltage and current density of the metal (semiconductor), and $R_{m}^{sh}$ and $R_{s}^{sh}=1/(eN_p \mu_{\rm cont})$ are the corresponding sheet resistance. 
Here, $N_p$ is the carrier density in the semiconductor, and $\mu_{\rm cont} $ is the hole mobility under the contact region, which can be different from the typical channel hole mobility $\mu$ (outside the contact region).

\begin{figure*}[t]
  \includegraphics[width=0.95\textwidth]{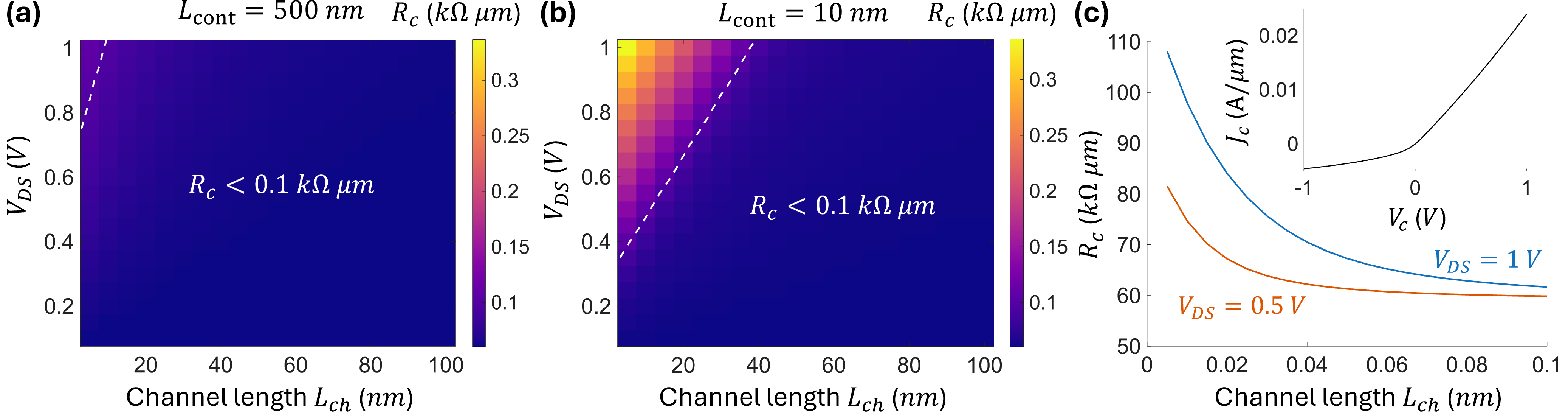}
  \caption{{\bf Contact resistance scaling as a function of channel length $L_{ch}$ and applied voltages $V_{DS}$.} (a, b) $R_c$ of Pd-WSe$_2$ \cite{Lance2026Science} as a function of $V_{\rm DS}$ and $L_{\rm ch}$ for contact lengths $L_{\rm cont}$=500 nm (a) and 10 nm (b) at $N_p =8\times 10^{12} \; {\rm cm^{-2}}$. Dashed white lines indicate the contour $R_c=0.1\, {\rm k\Omega\, \mu m}$. (c) $R_c$ increases as $L_{\rm ch}$ decreases. (inset) Current density $J_c$ as a function of voltage drop across the contact region $V_c$ shows saturation behaviour for large negative $V_c$. We set $L_{\rm cont}=500$ nm in (c). 
}
  \label{Fig2}
\end{figure*}

For a voltage drop $V_c$ across the contact length $L_{\rm cont}$ from metal to semiconductor, we implement the boundary conditions: $J_{s,0} =0$, $V_{m,0} = V_{c}$, $J_{m,L_{\rm cont}} = 0$ and $V_{s,L_{\rm cont}} = 0$. 
The current flowing through the contact is $J_c(V_c) = J_{m ,0}$.
Finally, the current flowing through the MSM structure for a source-drain voltage $V_{\rm DS}$ is obtained by self-consistently solving: 
\be\label{eq:j_msm}
J=J_c(V_{\rm DS}-V_1) = -J_c(-V_2) = \frac{\sigma_{\rm ch} (V_1-V_2)}{L_{\rm ch}},
\ee
where $V_1$ and $V_2$ are the voltages at either end of the channel (see Fig.~\ref{Fig1}a), $\sigma_{\rm ch} $ and $L_{\rm ch}$ are the channel conductivity and length. The contact resistance is defined as $R_c = (R_{\rm tot} - R_{\rm ch})/2$, where $R_{\rm tot} = V_{\rm DS}/J$ and $R_{\rm ch} = L_{\rm ch}/\sigma_{\rm ch} $ are the total and channel resistance respectively.

The edge contact between a metal and a 2D semiconductor can be modeled with a Schottky barrier at the edge's interface with height $\phi_S$. Here, we use a triangular barrier approximation with width $w(V)$, the in-plane tunneling probability of an electron at $k_z=0$ is:  
\be
T_e (\epsilon_{\vec k}) =  \min \left\{\exp\left[ -\frac{4t \sqrt{2m_0} w(V)}{3\hbar (\phi_i-V)} \left(\phi_S -\epsilon_{\rm inj}^{s} \right)^{3/2}\right], 1  \right\}
\ee
where $w(V) = \sqrt{2\varepsilon_s d_s (\phi_i - eV -k_BT)/(e^2 N_p)}$ is the Schottky barrier width, $\varepsilon_s$ is the semiconductor dielectric constant, $d_s$ is the thickness of the semiconductor, and $\phi_i = \phi_S -\epsilon_F+ \epsilon_{\rm VBM}$. 
Edge tunneling preserved transverse momentum, with $\eta = 0$ in Eq.~(\ref{eq:T_final}). 
The edge contact resistance in the MSM device is solved in a fashion similar to Eq.~(\ref{eq:j_msm}), with $L_{\rm ch}$ replaced by $L_{\rm ch} -  w(V_{\rm DS} - V_1) - w(-V_2)$ for the channel transport.
Here we have assumed that the metal electrodes (perfect conductor) have negligible resistance.

\addXY{\it Lowering vertical contact resistance for p-type 2D semiconductor.} We now examine the scaling law of p-contact resistance in MSM vertical configuration and identify the key factors to achieve ultralow $R_c$. 
We first analyse the scaling of $R_c$ as a function of semiconductor carrier density and compare our calculated results with experiments. 
In particular, we  consider the vertical contact configuration of metal-WSe$_2$ monolayer attached to a back gate (inset of Fig.~\ref{Fig1}e), which tunes the hole density $N_p$ in the p-type semiconducting WSe$_2$. 

Fig.~\ref{Fig1}e shows our numerical calculation of $R_c$ as a function of $N_p$ for a range of metal-monolayer WSe$_2$ vertical contact systems compared with recent experimental measurements (symbols)~\cite{Das2026-MoSe2-WSe2,Sun2026-NbWSe2, Zhang2026-SnS-WSe2, Lance2026Science}.  
To simulate the experiments of Ta-MoSe$_2$/Pd-WSe$_2$ and SnS/Au-WSe$_2$ contacts, we employed TLM approach to account for both the current flowing between MoSe$_2$ (SnS) and WSe$_2$ as well as current between Pd (Au) and MoSe$_2$ (SnS).
See {\bf SI} for detailed calculations and parameters used.
Here the channel length $L_{\rm ch}$ is ranged from 50 to 200 nm.

We see that $R_c$ of p-contact remains high compared to n-contact.
To further lower $R_c$, we propose a 2D topological semimetal AgSe on Ag substrate to be a new promising electrode material for WSe$_2$. 
Our first-principles calculation (Appendix C) predicts that the AgSe overlayer is stabilized by epitaxial interaction with the Ag substrate (to be validated by future experiment) and the heterostructure hosts highly hybridized metallic bands at the AgSe interface, forming a vdW contact with WSe$_2$. 
Due to the thin vdW barrier and large density of states in AgSe, $R_c$ approaches the 100 $\Omega \mu$m regime at $L_{\rm ch}=200$ nm and $N_p > 10^{13} \; {\rm cm^{-2}}$ (Fig.~\ref{Fig1}e orange solid line). 

Our calculations show that $\log R_c$ versus $\log N_p$ has a gradient of $\sim -1$ for all metallic contact materials (NbWSe$_2$, Ta-MoSe$_2$ and Pd), which is in agreement with experiments at high doping densities. 
This scaling suggests that the contact current is source-limited by the semiconductor carrier density. 
For p-contact and $V_c<0$, the semiconductor valence band maximum (VBM) is shifted downward, and the current is limited by the hole density around $\epsilon_{\rm VBM}$ available for electron tunneling: $J_v (V_c) \sim \mathcal T(\epsilon_{\rm VBM}, V_c) v_{\rm inj} N_p $, where $v_{\rm inj} \approx \sqrt{2(\epsilon_{\rm VBM} +V_c-\epsilon_n^0)/m_{m}}$ is the electron injection velocity. 
For non-metallic electrode like SnS/Au, however, the $\log R_c$ versus $\log N_p$ plot has a much gentler gradient. This is because the contact current is source-limited by the carrier density of the semiconducting SnS and less sensitive to the carrier density in WSe$_2$. 

Interface engineering to reduce the vdW barrier height and width is critical to achieve ultralow $R_c$. For $\phi_r \gg V$, the contact current $J_c \sim e^{-t\sqrt{\phi_r}}$ increases exponentially with $t\sqrt{\phi_r}$. In Fig.~\ref{Fig1}e, we show that reducing the vdW barrier to zero can push the AgSe/Ag-WSe$_2$ contact resistance (orange dashed line) to approach the quantum limit of p-contact resistance (black dashed line). 

Mobility degradation of 2D semiconductors in contact with metal is undesired, leading to higher $R_c$. Under the contact region, the semiconductor carrier mobility $\mu_{\rm cont}$ is lower than its intrinsic value $\mu$ due to interfacial chemical reactions and disorder \cite{Schauble2020,Smyth2016,Smyth2017}, leading to increasing $R_s^{sh}$ and $R_c$. 
As an illustration, we plot $R_c$ for the NbWSe$_2$-WSe$_2$ contact and find that $R_c \propto 1/\sqrt{\mu_{\rm cont}/\mu} \propto \sqrt{R_{sh}^s}$ (Fig.~\ref{Fig1}f), which is consistent with prior findings \cite{schroder2015}. 

\addXY{\it Length and $V_{\rm DS}$ scaling of the p-contact resistance.} Down-scaling the channel length to sub-10 nm regime while maintaining low contact resistance is the central challenge in developing next generation FETs. 
Fig. ~\ref{Fig2} shows our calculated $R_c$ as a function of $V_{\rm DS}$ and $L_{\rm ch}$ for Pd-WSe$_2$ \cite{Lance2026Science}. 
The scaling of $R_c$ as a function of $L_{\rm ch}$ depends on $V_{\rm DS}$. 
At small $V_{\rm DS}$, $R_c$ is relatively insensitive to $L_{\rm ch}$. As $V_{\rm DS}$ increases, the contact resistance generally increases. 
Furthermore, $R_c $ increases as $L_{\rm ch}$ decreases at short channel length and large $V_{\rm DS}$. The white dashed lines in Fig.~\ref{Fig2}a and b indicate $V_{\rm DS}$ when $R_c = 100 \,\Omega \, {\rm \mu m}$ for each $L_{\rm ch}$, which is linear with gradient $J/\sigma_{\rm ch}$, where $J$ is the current density when $R_c=100 \, {\rm \Omega \, \mu m}$. 

\begin{figure}
  \includegraphics[width=\columnwidth]{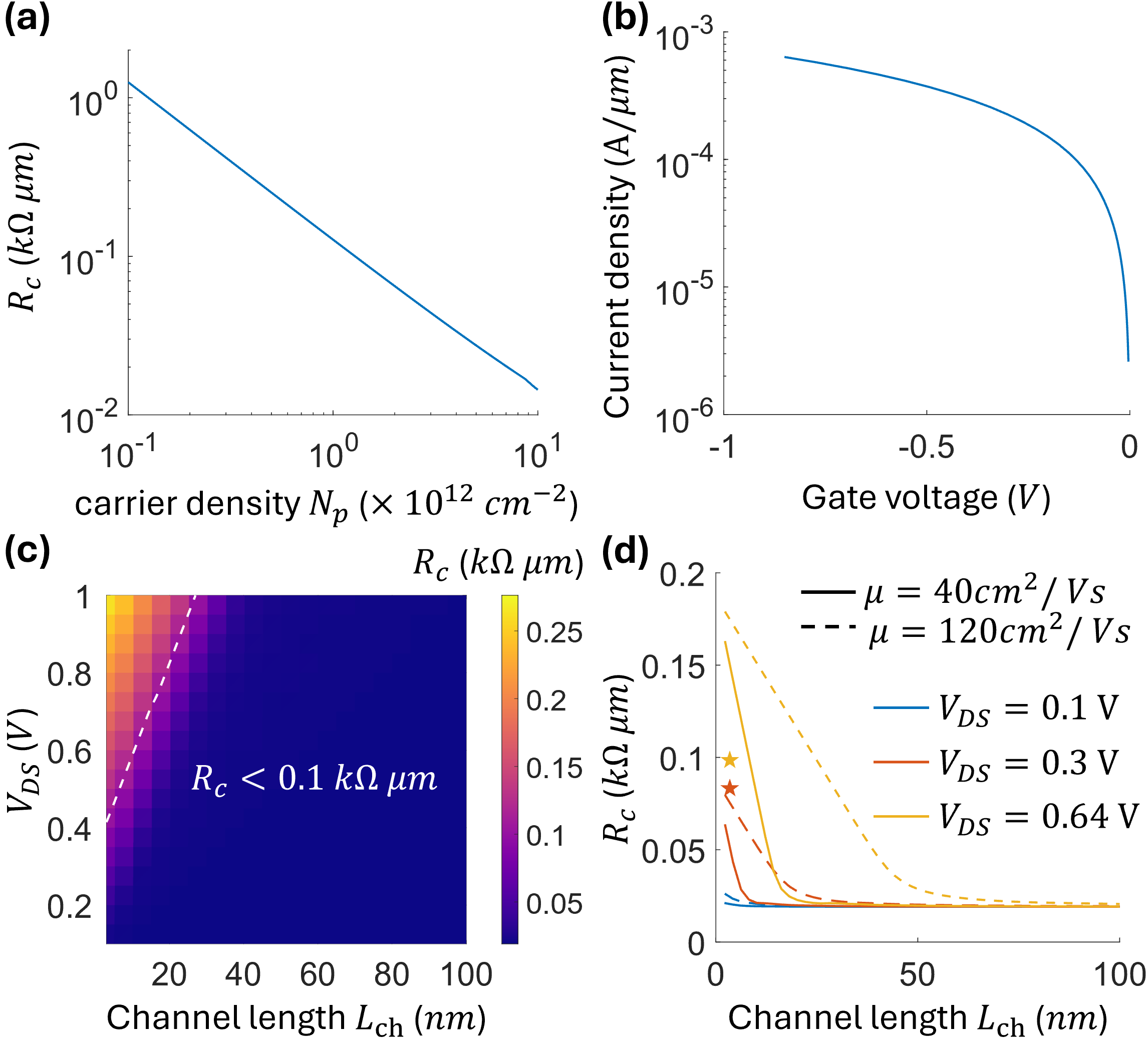}
  \caption{{\bf Contact resistance for CuS-WSe$_2$ system.} (a) Contact resistance $R_c$ as a function of carrier density. (b) Current density versus gate voltage. We have set $\phi_S = 10$ meV, $V_{\rm DS} = 0.64$ V, $L_{\rm ch} = 100$ nm, $\mu=60 \; {\rm cm^2/Vs}$, and channel gate capacitance $0.02 \; {\rm F/m^2}$ for a 10 nm thick HfO$_2$ top gate. (c) $R_c$ increases for larger $V_{\rm DS}$ and shorter $L_{\rm ch}$. White dashed line indicates $R_c=0.1 \,{\rm k\Omega \,\mu m}$. (d) $R_c$ increases linearly as $L_{\rm ch}$ decreases at small $L_{\rm ch}$. Markers denote first-principle quantum atomistic calculations at 3.4 nm. For (c) and (d), we set to $N_p\approx 7.3 \times 10^{12} \; {\rm cm^{-2}}$. 
}
  \label{Fig3}
\end{figure}

To understand the scaling of $R_c$ at small $L_{\rm ch} <$ 100 nm, we plot the contact current as a function of potential drop $V_c$ across the contact length $L_{\rm cont}$ in Fig.~\ref{Fig2}c inset. 
We find that the current becomes source-limited at large $|V_c|$. 
In particular, 
for $V_c < \epsilon_F - \epsilon_{\rm VBM}$, the current is limited by the small number of holes around $\epsilon_{\rm VBM}$, which is shifted downward by $V_c$. 
The current saturates as $V_c$ becomes more negative as the effective potential barrier for electrons increases, which explains the increment of contact resistance $J_c/V_c$ for larger voltages. 

For a fixed $V_{\rm DS}$, as $L_{\rm ch}$ decreases, we have smaller channel resistance $R_{\rm ch}$ leading to a decrease in the voltage drop across the channel and an increase in the voltage across the contacts. 
Thus, for small $L_{\rm ch}$ and large $V_{\rm DS}$, the current will approach the saturation regime, and $R_c$ exhibits an approximately linear inverse dependence on channel length,
as shown by the blue curve in Fig.~\ref{Fig2}c. 
As expected in Fig.~\ref{Fig3}d, such current saturation and $R_c$ scaling also manifest in the edge contact configuration. 

Comparing Figs.~\ref{Fig2}a and ~\ref{Fig2}b,
this effect of current saturation will increase $R_c$ about 3X for smaller contact length of $L_{\rm cont}$= 10 nm from a recent experiment of $L_{\rm cont} = 500$ nm \cite{Lance2026Science}, which will pose a limit to down-scaling of 2D FET to sub-10 nm to maintain low contact resistance using the same electrode.
This limitation is due to small transfer length $L_T \sim 5$ nm for Pd caused by current saturation at $V_c<0$. 
Similar scaling and limitation are also shown in other electrode materials such as NbWSe$_2$, see {\bf SI}.  
In Table~\ref{tab:LT} in Appendix B, we compare $L_T$ of several contact materials and show good agreement with direct experimental measurement using XSTM~\cite{Yang2026}. We identify Bi and Pd for n- and p-contact materials with the shortest transfer lengths owing to small $R_c$. 
Due to fabrication challenges, there are NO experimental measurement of the p-contact resistance at small $L_{\rm ch}$. 
To our best knowledge, the shortest channel length that has been fabricated for p-contact has reported $R_c =145 \; {\rm \Omega \,\mu m}$ at $L_{\rm ch} = 50$ nm~\cite{Lance2026Science}, see Fig.~\ref{Fig1}e.
It will be of interest to conduct more experiments of sub-50 nm channel length to confirm the limit of contact resistance due to the effect of current saturation as reported in this paper.

\addXY{\it Ultralow p-contact resistance for CuS-WSe$_2$ heterostructure.} We show that a new high-work-function 2D semimetal CuS can also be a promising contact material for WSe$_2$ to reach $R_c<$ 100 $\Omega {\rm \mu m}$ contact resistance.
By stacking with WSe$_2$, electronic bands in the CuS-WSe$_2$ hetero-bilayer become highly hybridized between CuS and WSe$_2$, and electron wavefunctions near the Fermi surface are delocalized across the layers (see {\bf SI}).
Thus, the CuS-WSe$_2$ hetero-bilayer can be treated as a 2D metal sheet, and the main contribution to the contact resistance is accounted by the edge contact model between the proposed metallic CuS-WSe$_2$ electrode and the semiconducting WSe$_2$ channel. 

We consider CuS-WSe$_2$-CuS structure with a top gate in the channel region to tune the channel hole density $N_p$. DFT calculation shows no observable Schottky barrier at the contact. 
Figures.~\ref{Fig3}a and b show the p-contact resistance and current density for a small $\phi_S =10$ meV and $L_{\rm ch} =100$ nm. We find that the p-contact resistance is significantly reduced down to the $R_c<$ 100 $\Omega \; {\rm \mu m}$ range for moderate $N_p > 10^{12} {\rm \; cm^{-2}}$ with current density approaching the order of $10^{-3}\;  {\rm A/\mu  m}$ at $V_{\rm DS} = 0.64$ V. 
  
Similar to vertical contact, edge contact resistance also increases with increasing $V_{\rm DS}$ and decreasing $L_{\rm ch}$, as shown in Fig.~\ref{Fig3}c and d. 
For the ultra-nanoscaled $L_{\rm ch}<$ 10 nm and large $V_{\rm DS}$, $R_c$ for CuS-WSe$_2$ contact is still lower than most known contact electrodes reported in prior experiments (Fig.~\ref{Fig1}e).
As shown in Fig.~\ref{Fig3}d, $R_c$ increases linearly as $L_{\rm ch}$ decreases in the short channel limit like sub-10 to sub-50 nm depending on $V_{DS}$= 0.1 to 0.64 V.
This is due to thicker Schottky barrier and current saturation at large positive $V_c$. 
Our prediction of increasing $R_c$ with $V_{\rm DS}$ qualitatively agrees with full quantum atomistic calculations (stars in Fig.~\ref{Fig3}d, see {\bf SI} for details) at $L_{\rm ch}$ = 3.4 nm. 
While a full quantum treatment is required to accurately model the electron transport at sub-5 nm, we believe that the general trend of increasing $R_c $ with decreasing $L_{\rm ch}$ should still hold due to current saturation at large $V_c$ regardless of the details of transport in the channel and interface region.

In summary, our paper provides a physics-consistent and yet simple model for the MSM contact resistance for a channel length of 5 to 100s nm. 
We analyze the scaling of $R_c$ as a function of key experimentally relevant parameters and outline pathways for further reducing $R_c$ toward the quantum limit. 
Two new materials as electrodes have been suggested to further reduce the p-contact resistance to $R_c<$ 1000 and $R_c<$ 100 $\Omega \; {\rm \mu m}$, respectively. Searching for new electrode materials that are thermally stable at room temperature with low tunneling barrier is essential to achieve ultralow $R_c$ by avoiding the effect of source-limited saturation identified in this paper especially at sub-50 nm scale.  
Our formulation can be readily incorporated into the current TCAD models for development of 2D FETs in sub-50 nm regime to be verified with future experiments.
The model is also valid for n-type contact as discussed in {\bf SI} and Appendix A. We note that the interfacial physics discussed here is also critical for AC contact resistance with skin effect modulation ~\cite{Faisal2026}, which is subject to future studies. 

{\it Acknowledgments.} This research is funded by the Singapore A*STAR IRG grant (M23M6c0102). Y. S. A. is supported by the Singapore National Research Foundation (NRF) Frontier Science Competitive Research Programme (F-CRP) under the award number NRF-F-CRP-2024-0001 and the Kwan Im Thong Hood Cho Temple Early Career Chair Professorship.

\bibliography{refs}

\onecolumngrid

\smallskip

\begin{center}
\textbf{\large End Matter}
\end{center}

\twocolumngrid

\textcolor{blue}{\it Appendix A: contact resistance of n-type 2D semiconductors.} We show that our models can be readily adapted for contact resistance of n-type semiconductors. For n-type contact, electrons tunnel into the conduction band of the semiconductor $\epsilon_{\vec q} = \epsilon_{\rm CBM} + \frac{\hbar^2 q^2}{2m_s} $, where $\epsilon_{\rm CBM}$ is the conduction band minimum (CBM), and $m_s>0$. It follows that the tunneling probability in Eq.~(\ref{eq:T_final}) is replaced with (see {\bf SI} for derivation)
\be
\mathcal T (\epsilon_{n, \vec k}, V) = T_v (\epsilon_{n, \vec k}) \big[\eta + (1-\eta) \Theta(  \epsilon_{\rm inj}^s+ V -\epsilon^s_{\rm CBM} ) \big].
\ee
Again, $\eta =0$ implements the conservation of transverse momentum and $\eta = 1$ for the case when the transverse momentum is not conserved.

\begin{figure}
    \centering
    \includegraphics[width=\linewidth]{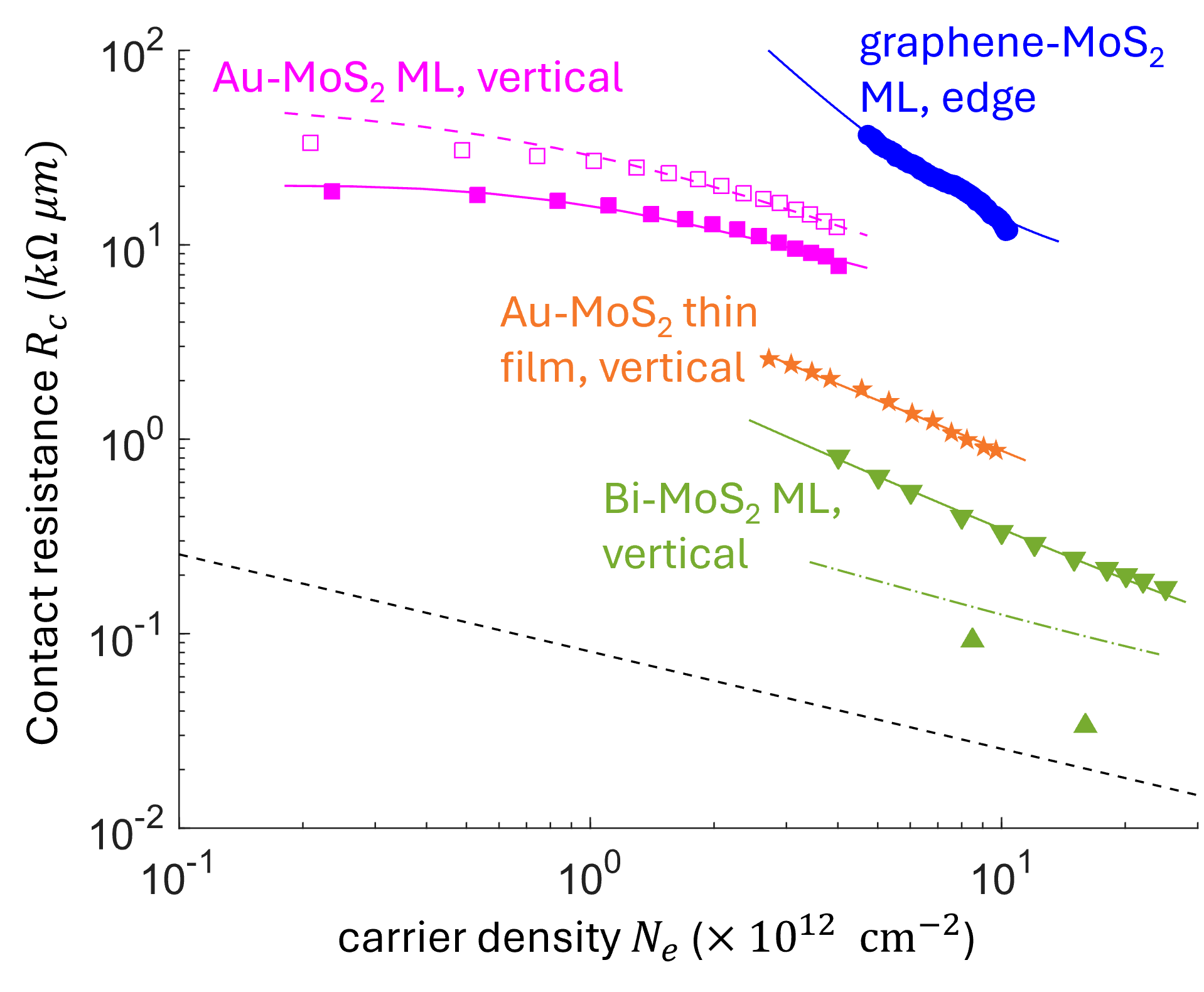}
    \caption{Contact resistance of n-type 2D semiconductor as a function of semiconductor electron density $N_e$. Filled (hollow) markers denotes the experimental measurements of $R_c$ at $V_{\rm DS} = 1$ V ($V_{\rm DS} = 0.1$). Solid (dashed) lines are by using our model with $\eta = 0$ at $V_{\rm DS} =1$ V ($V_{\rm DS} =0.1$ V). The dash-dotted line is obtained by taking $\eta = 0$, $V_{\rm DS} = 1$ V, $\phi_r = 0$ and $t=0$, which indicates the lowest $R_c$ for Bi-MoS$_2$ ML contact predicted by our theory. In our calculations, $L_{\rm ch}$ for all vertical contacts is set as the shortest channel length used in the respective experiments. For Bi-MoS$_2$, $L_{\rm ch} = 50$ nm for Ref.~\cite{Lance2026Science} (upright green triangles) and $L_{\rm ch} = 100$ nm for Ref.~\cite{Wu2026} (inverted greeen triangles); for Au-MoS$_2$ ML $L_{\rm ch} = 200$ nm~\cite{Smithe2017}; for Au-MoS$_2$ thin film $L_{\rm ch} = 100$ nm~\cite{English2016}. 
    For graphene-MoS$_2$ ML edge contact~\cite{Guimaraes2016}, $L_{\rm ch} =1.11 \;\mu$m is used. Other parameters are stated in the {\bf SI}. Black dashed line indicates the quantum limit.}
    \label{FigS1}
\end{figure}

In Fig.~\ref{FigS1}, we compare our model with the experimental data (markers) of $R_c$ for a range of contact materials in vertical and edge contact configurations. The solid and dashed lines corresponds to $\vec k_{\rm tr}$-conserving case $\eta = 0$ at $V_{\rm DS} =1$ V and $V_{\rm DS} = 0.1$ V respectively. The dash-dotted line is obtained for $\vec k_{tr} $-nonconserving case with $\phi_r=0$ and $t=0$, which indicates the lowest $R_c$ for Bi-MoS$_2$ vertical contact. Details of parameters can be found in the {\bf SI}. Our model shows excellent agreement with the experimental data except for the ultralow $R_c$ of Bi-MoS$_2$ monolayer vertical contact (upright triangles) in Ref.~\cite{Lance2026Science}, which is lower than our predicted minimum $R_c$. One potential reason could be the possibly higher doping of the Bi semimetal in experiment, which leads to lower electrode sheet resistance and higher current. Inelastic scattering of electrons at the interface can relax the energy conservation requirement and enlarge the phase space allowed for the transmission, further lowering the contact resistance. Another possible reason is that we did not include the metallic electrode in contact with Bi in our model, which may further lower the measured contact resistance due to its small sheet resistance. 

\begin{table*}
\centering
\begin{tabular}{c|c|c|c|c|c}
\hline
\hline
 & $N_{s(p)}$ ($10^{12} \; \rm{cm^{-2}}$) & simulated $L_{T1} $(nm)& simulated $L_{T2} $ (nm)& experimental (nm)& Ref. \\ \hline
NbWSe$_2$-WSe$_2$ ML & 4.8 & 62 & 120 & 90 & TML expt~\cite{Sun2026-NbWSe2} \\
Pd-WSe$_2$ ML & 7.5 & 4.4 & 4.6 & 7.7 & TML expt.~\cite{Lance2026Science}\\ 
Au-MoS$_2$ ML & 3.9 & 148 & 136 & 94 & TML expt.~\cite{Smithe2017} \\
Au-MoS$_2$ thin film & 9.5 & 66 & 43 & 42 & TML expt.~\cite{English2016} \\
Bi-MoS$_2$ ML & 6.8 & 2.8 & 2.8 & $L_{T1}=2.4$ , $L_{T2}= 2.9 $ & XSTM expt.~\cite{Yang2026} \\
\hline
\hline
\end{tabular}
\caption{Comparison of our simulated transfer length $L_T$ with experimental measurement for selected p- and n-type contacts of using different monolayer (ML) 2D semiconductors.
The applied voltage for all calculations is set at $V_{\rm DS} = 1$ V. }
\label{tab:LT}
\end{table*}

\textcolor{blue}{\it Appendix B: Transfer length for vertical contact.}
Reducing the contact length $L_{\rm cont}$ is crucial for the down-scaling of 2D FET. The contact resistance will increase significantly when $L_{\rm cont}$ approaches the transfer length $L_T$.
In this section, we calculate $L_T$ at contact 1 and 2 for several n- and p-type contacts and compare them with experiments, see Table~\ref{tab:LT}. Here, the transfer length is defined as the length for the voltage $V_{c,x} = V_{m,x} -V_{s,x}$ between the metal electrode and the semiconductor to decrease by $1/e$. To calculate $L_T$, we fit the voltage away from the physical edge of the electrode towards the channel side using an exponential decay $V_{c,x} =  e^{(x-x_0)/L_T}$, and extract $L_{T1}$ at contact 1 and $L_{T2}$ at contact 2 (see Fig.~\ref{Fig1}a). For NbWSe$_2$-WSe$_2$ ML, Pd-WSe$_2$ ML, Au-MoS$_2$ ML and Au-MoS$_2$ thin film, we compare our simulated transfer length (see {\bf SI} for details of parameters used) with the extracted values from TML measurements. For Bi-MoS$_2$ ML, we compare our simulation with direct measurement of transfer length using cross-sectional scanning tunnelling microscopy (XSTM)~\cite{Yang2026}. To compare with Ref.~\cite{Yang2026}, we have used $\phi_r=0$, $t=0$, $\eta = 1$, $L_{\rm cont1} = L_{\rm cont2} = 200$ nm, $L_{\rm ch} = 2 \;\mu$m and $V_{\rm DS}  =1$ V. The MoS$_2$ mobility $\mu \approx 12 \; {\rm cm^2/Vs}$ is extracted from the TML measurement in Ref.~\cite{Yang2026}.

We identify Bi and Pd as n- and p-type contact materials with the shortest transfer length due to their small contact resistivity. While our simulated transfer lengths show varying degree of agreement with the TML experiments, our $L_T$ agrees closely with the direct measurement of transfer length using XSTM for Bi contact~\cite{Yang2026}. Furthermore, for Bi and Pd, the transfer lengths at contact 1 ($V_c>0$) and contact 2 ($V_c<0$) are almost equal. This is due to the small $R_c$ and thus the small magnitude of $V_c$ so that the contact is almost Ohmic. For other contacts materials, we generally have $L_{T1} \neq L_{T2}$ due to complicated $V$-dependent contact resistivity.

\textcolor{blue}{\it Appendix C: First-Principles calculations of WSe$_2$/AgSe/Ag heterostructure.} First-principles calculations were performed within the framework of density functional theory (DFT) using the Vienna Ab initio Simulation Package (VASP)~\cite{Kresse1993, Kresse1996}. The electron–ion interactions were described by the projector augmented-wave (PAW) method~\cite{Blochl1994, Kresse1999}, and the exchange–correlation functional was treated using the generalized gradient approximation in the Perdew Burke Ernzerhof (PBE) form~\cite{Perdew1996}. A plane-wave cutoff energy of 500 eV was adopted for all calculations. The atomic positions were fully relaxed until the total energy and atomic forces converged to $10^{-6}$ eV and $10^{-3}$ eV ${\rm \mathring{A}^{-1}}$, respectively. The Brillouin zone was sampled using a $\Gamma$-centered $3\times3\times 1$ $k$-point mesh~\cite{Monkhorst1976}. The van der Waals interaction was considered using the DFT-D3 method with Becke–Johnson damping~\cite{Grimme2010,Grimme2011}. Owing to the asymmetric slab geometry, a dipole correction was applied along the out-of-plane direction~\cite{Neugebauer1992}. A vacuum layer of approximately 20 ${\rm \mathring A}$ was introduced along the out-of-plane direction to avoid artificial interactions between periodic images.

\begin{figure}
    \centering
    \includegraphics[width=\linewidth]{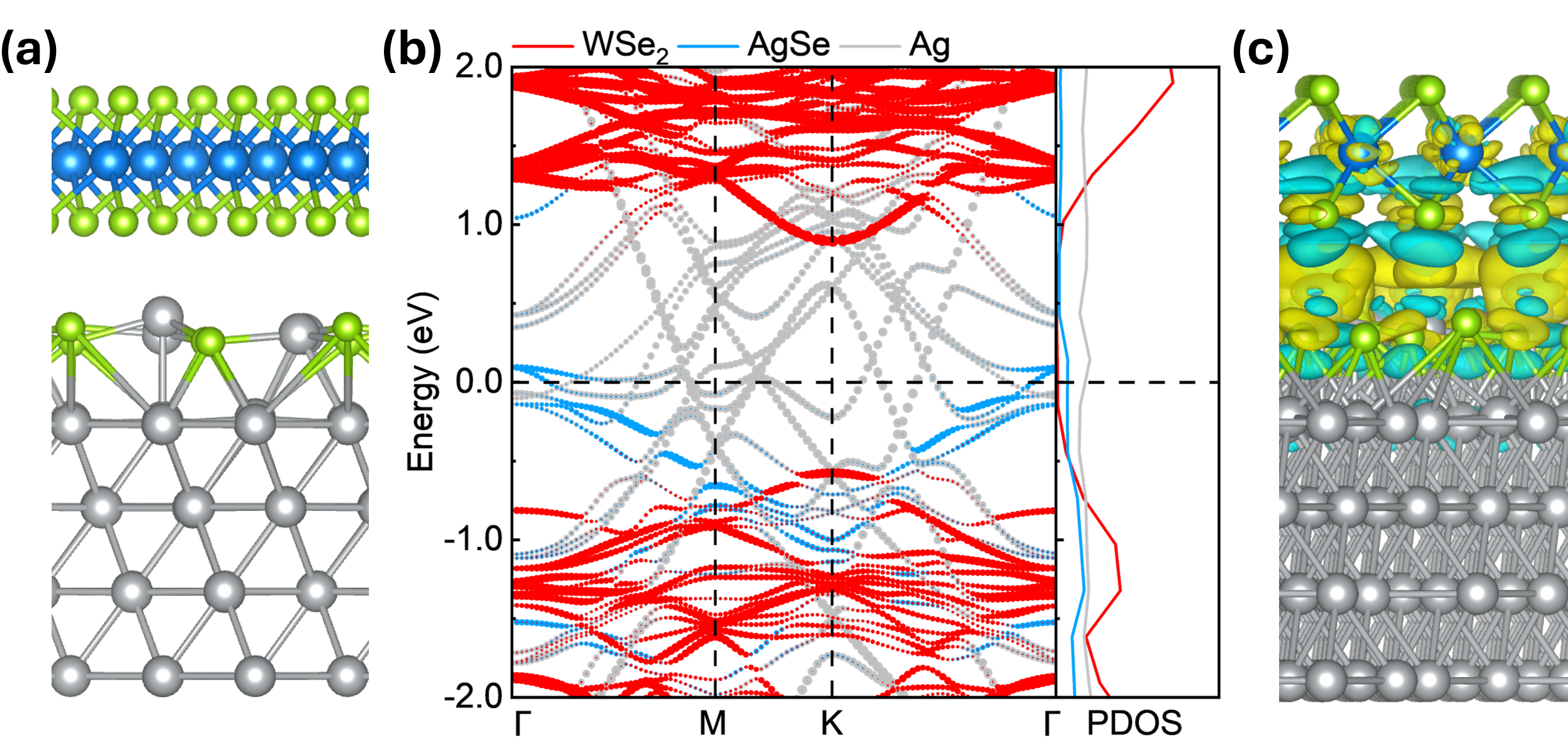}
    \caption{Atomic and electronic structures of the WSe$_2$/AgSe/Ag heterostructure. (a) Side view of the optimized atomic configuration. (b) Projected band structure and corresponding projected density of state, where the contributions from WSe$_2$, AgSe, and Ag are marked in red, blue, and gray, respectively. The horizontal dashed line denotes the Fermi level. (c) Charge density difference of the heterostructure, showing charge accumulation (yellow area) and depletion (blue area) at the interface. }
    \label{FigS3}
\end{figure}

As shown in Fig.~\ref{FigS3}, the introduction of the AgSe interfacial layer leads to a nontrivial interfacial coupling between WSe$_2$ and the Ag electrode, rather than a simple physical contact. The projected band structure indicates that the electronic states near the Fermi level are mainly contributed by the metallic Ag/AgSe components, while the intrinsic band features of WSe$_2$ are largely preserved. This suggests that the AgSe layer can partially suppress the strong orbital hybridization between WSe$_2$ and the Ag substrate. Moreover, the charge density difference (as shown in Fig.~\ref{FigS3}c) reveals a pronounced electron transfer from Ag as well as WSe$_2$ into the AgSe layer, This results in a high density of state at AgSe interface around the Fermi surface and hole injection in WSe$_2$, contributing to the relatively low contact resistance as shown in Fig.~\ref{Fig1}e.

\end{document}